\documentclass[%
 reprint,
superscriptaddress,
 amsmath,amssymb,
 aps,
 prl,
]{revtex4-2}

\usepackage{graphicx}% Include figure files
\usepackage{dcolumn}% Align table columns on decimal point
\usepackage{bm}% bold math
\usepackage[colorlinks=true, citecolor=blue, linkcolor=blue, urlcolor=blue]{hyperref}

\begin{document}

\preprint{APS/123-QED}

\title{
Aharonov–Bohm interference from coherent spin-polarized edge transport in Fe(Te,Se) superconducting rings
}

\author{Mohammad Javadi Balakan}
\email{mjavadibalakan@albany.edu}
\affiliation{College of Nanotechnology, Science, and Engineering, State University of New York, Albany, NY 12203, USA.}

\author{Shiva Heidari}
\affiliation{Physics Department, City College of the City University of New York, NY 10031, USA.}

\author{Genda Gu}
\affiliation{Condensed Matter Physics and Materials Science Division, Brookhaven National Laboratory, Upton, NY 11973, USA.}

\author{Qiang Li}
\affiliation{Condensed Matter Physics and Materials Science Division, Brookhaven National Laboratory, Upton, NY 11973, USA.}
\affiliation{Department of Physics and Astronomy, Stony Brook University; Stony Brook, New York 11794, USA.}

\author{Kenji Watanabe}
\affiliation{National Institute for Materials Science, 1-1 Namiki, Tsukuba 305-0044, Japan.}

\author{Takashi Taniguchi}
\affiliation{National Institute for Materials Science, 1-1 Namiki, Tsukuba 305-0044, Japan.}

\author{Ji Ung Lee}
\email{jlee1@albany.edu}
\affiliation{College of Nanotechnology, Science, and Engineering, State University of New York, Albany, NY 12203, USA.}

\date{\today}
\begin{abstract}
We report the coexistence of Aharonov–Bohm and Little–Parks oscillations in mesoscopic Fe(Te,Se) rings. The magnetoresistance shows two distinct periodicities: an $h/e$ component from ballistic edge interference and an $h/2e$ component from fluxoid quantization of Cooper pairs. Aharonov–Bohm oscillations persist deep into the superconducting phase, exhibit current–field symmetry, and follow a temperature dependence captured by a helical Luttinger liquid model, consistent with edge states in a topological superconductor.
\end{abstract}

\maketitle

Quantum electronics relies fundamentally on the phase coherence of electronic wavefunctions enabling interference-based functionalities. In normal metals, such coherence gives rise to Aharonov–Bohm (AB) interference of single electrons with periodicity $\Phi_e = h/e$ \cite{ab1959, washburn1986}. In superconductors, phase coherence manifests through flux quantization in units of $\Phi_{2e} = h/2e$, reflecting the Cooper-pair charge and giving rise to Little–Parks (LP) oscillations in mesoscopic rings \cite{littlepark1962, moshchalkov1995}. While both effects originate from quantum phase winding in closed-loop geometries, they arise from fundamentally different transport mechanisms, single-particle versus collective, and typically appear in distinct physical regimes \cite{aronov1987}. Their coexistence becomes particularly intriguing in systems hosting topological superconductivity, where a gapped bulk coexists with symmetry-protected edge states that support coherent single-electron transport \cite{read2000, alicea2012, holst2022}. These edge channels are predicted to host exotic quasiparticles such as Majorana zero modes with non-Abelian statistics, which form the foundation for fault-tolerant quantum computation \cite{ivanov2001, kitaev2002, nayak2008}. Experimentally, however, simultaneous observation of macroscopic phase coherence and localized edge conduction remains rare, likely due to the dominance of bulk superconducting carriers.

The iron-based superconductor Fe(Te,Se) has emerged as a compelling platform for intrinsic topological superconductivity. Prior studies have reported Dirac surface states \cite{wang2015, zhang2018, zhang2019} and zero-bias conductance peaks in vortex cores, interpreted as signatures of Majorana zero modes \cite{wang2018, zhu2020, machida2019}. More recently, we identified half-quantum vorticity in mesoscopic Fe(Te,Se) devices, revealing the presence of topological defects essential for Majorana physics. In addition, our transport measurements uncovered spin-polarized, symmetry-sensitive quantum oscillations, providing direct evidence for the nontrivial topology of the superconducting order parameter \cite{balakan2025observation}.

Here, we report the coexistence of AB and LP oscillations in mesoscopic Fe(Te,Se) rings, providing direct evidence for the simultaneous emergence of coherent edge states and Cooper pairs in a superconducting platform. The AB oscillations exhibit $\Phi_e$ periodicity and are confined to the ring edge, while the LP oscillations exhibit $\Phi_{2e}$ periodicity, reflecting fluxoid quantization through the effective ring area. Remarkably, the AB interference persists deep into the superconducting state and emerges only under finite DC bias, with a strong current–field symmetry and a temperature dependence indicative of ballistic transport. These features point to spin-polarized, phase-coherent edge conduction—a hallmark of topological superconductivity. Our findings establish Fe(Te,Se) as a unique platform where collective and single-particle coherence coexist, enabling new avenues to probe and control edge modes in superconducting quantum materials.

\begin{figure}[t]
\includegraphics[width=0.47\textwidth]{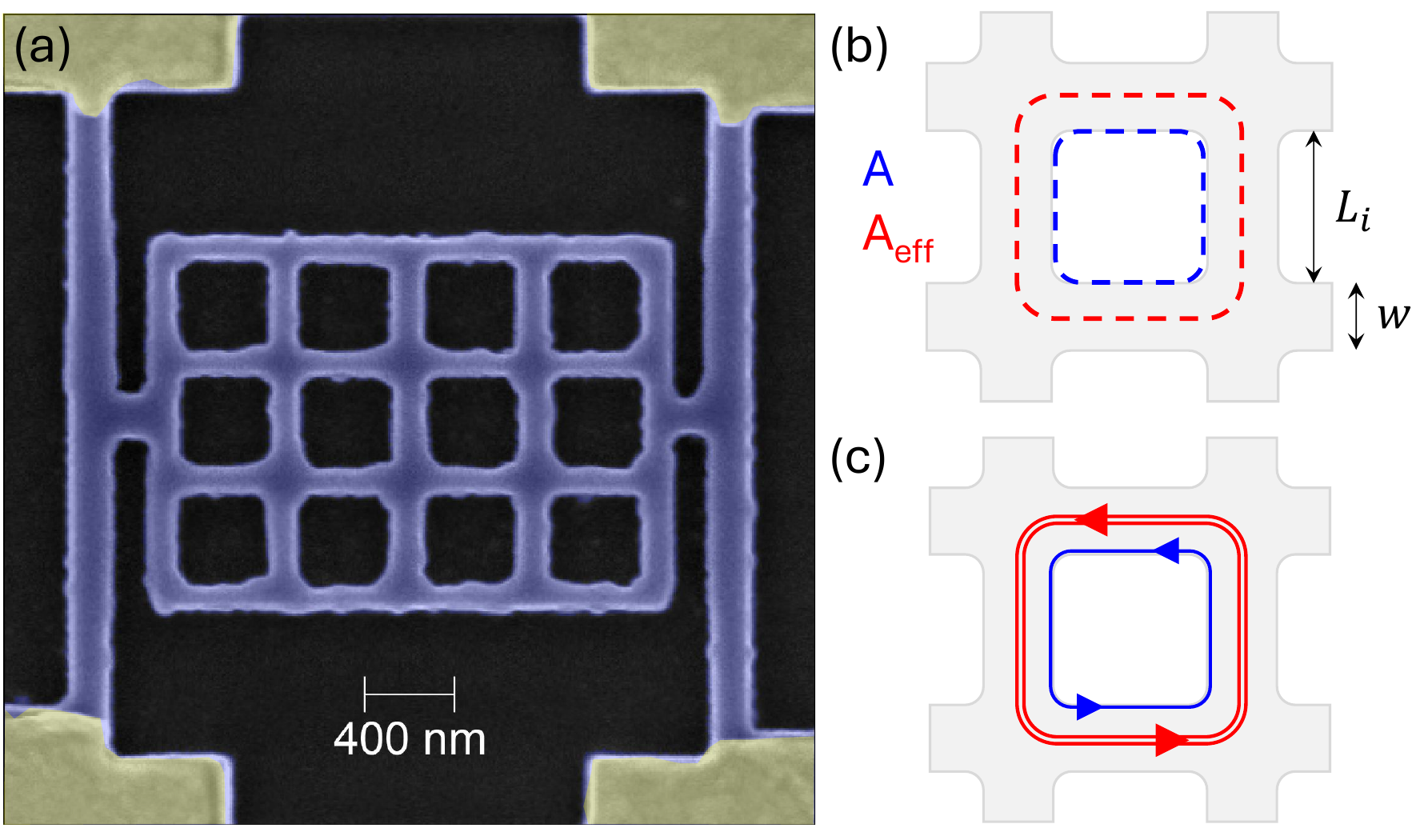}
\caption{\label{fig:sem_schem} (a) False-color scanning electron microscopy image of the device. The purple and yellow regions correspond to hBN/Fe(Te,Se) and Au/Ti/Fe(Te,Se), respectively. (b) Schematic of a single ring, indicating the inner side length $L_i$ and wall width $w$. (c) Schematic paths of circulating current along the inner edge (blue) and through the ring bulk (red), enclosing flux areas $A$ and $A_\mathrm{eff}$, respectively.}
\end{figure}

Figure~\ref{fig:sem_schem}(a) shows a false-color scanning electron micrograph of the device, consisting of an array of mesoscopic Fe(Te$_{0.55}$,Se$_{0.45}$) rings fully encapsulated in a thin hBN layer to prevent degradation. Electrical contacts to the Fe(Te,Se) layer are made via etched vias using Au/Ti electrodes. Transport measurements were performed using a four-terminal AC+DC lock-in technique with a 1~$\mu$A AC excitation and a variable DC offset. Details of the fabrication and measurement setup are provided in Ref.~\cite{balakan2025observation}. Each ring features an average inner side length of $L_i = 406 \pm 19$~nm and wall width $w = 118 \pm 17$~nm [Fig.~\ref{fig:sem_schem}(b)]. The effective area relevant to magnetic flux depends on the nature and location of the circulating current. For Cooper pairs, fluxoid quantization is imposed on the macroscopic phase of the superconducting order parameter, leading to LP oscillations. Near the superconducting transition temperature, the London penetration depth exceeds the wall width, causing the supercurrent to flow along the midline of the ring wall. The corresponding enclosed area is given by $A_\text{eff} = L^2 \left(1 + (\text{w}/L)^2\right)$, where $L = L_i + \text{w}$, accounting for the midpoint of the superconducting wall with a geometric correction \cite{Groff1968}. A similar effective (or average) area is also used when studying AB interference in metallic or semiconductor rings, where phase-coherent trajectories extend across the ring width. In the case where the circulating currents are sharply confined to the inner edge, the enclosed area is purely geometric: $A = L_i^2$. We will refer to $A_\text{eff}$ and $A$ as the ``bulk'' and ``edge'' paths, respectively [Fig.~\ref{fig:sem_schem}(b,c)]. For each path, quantum oscillations may arise either from Cooper pairs ($\Phi_{2e}$) or from single electrons ($\Phi_e$), yielding four characteristic magnetic field scales as shown in Table~\ref{tab:table1}.

\begin{table}[t]
\caption{\label{tab:table1} Characteristic magnetic field periods for AB interference and LP fluxoid quantization. The edge and bulk paths enclose the inner area ($A$) and effective area ($A_\text{eff}$), respectively. Periodicities are defined as $\Delta B_{e(2e)} = \Phi_{e(2e)}/\text{Area}$.}
\begin{ruledtabular}
\begin{tabular}{ccc}
Area & $\Delta B_{e}$ (G)& $\Delta B_{2e}$ (G)\\
\colrule
\\
$A$ & 252$\pm$24 &  125$\pm$12 \\
(edge path) & (edge AB) & (edge LP) \\
\\
$A_\text{eff}$ &  143$\pm$13 &  72$\pm$7 \\
(bulk path) & (bulk AB) & (bulk LP)\\
\end{tabular}
\end{ruledtabular}
\end{table}

\begin{figure*}%[b]
\includegraphics[width=0.97\textwidth]{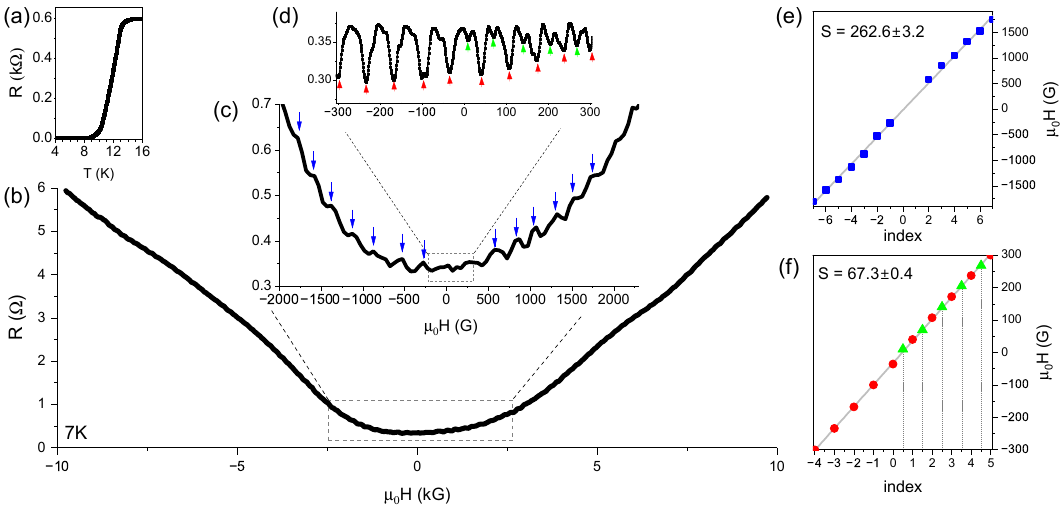}
\caption{\label{fig:MR7K} (a) Resistance versus temperature near the superconducting transition. (b) Magnetoresistance under coarse magnetic field sweep at 7~K and $I_\mathrm{DC} = 2.2$~$\mu$A. (c) Zoomed-in view of the low-field region in panel b, showing periodic oscillations marked by blue arrows. 
(d) MR at low magnetic fields, exhibiting two sets of periodic oscillations indicated by red and green arrows. (e) Magnetic field versus index from panel c, with a slope corresponding to AB oscillations along the ring edge. (f) Magnetic field versus index from panel d, with a slope consistent with LP oscillations from circulating bulk supercurrents. Red and green points correspond to full- and half-quantum vorticity, respectively. Linear fits (gray lines) are overlaid in panels e and f.}
\end{figure*}

The superconductivity transition is shown in Fig.~\ref{fig:MR7K}(a), with an onset near 14~K and a fully developed non-resistive phase below $T_c=8~K$. Figures~\ref{fig:MR7K}(b--d) represent magnetoresistance (MR) response under coarse and fine magnetic field sweeps collected at 7~K. At this temperature, the device resides in the superconducting state, with no measurable resistance. Applying a finite DC offset, superimposed on a 1~$\mu$A AC excitation, drives the device into a low-resistance regime, enabling measurement of voltage across the sample. Under a coarse magnetic field sweep from $-10$~kG to $+10$~kG [Fig.~\ref{fig:MR7K}(b)], the MR exhibits a parabolic trend due to Meissner screening. A zoomed-in view near zero field [Fig.~\ref{fig:MR7K}(c)] reveals periodic oscillations (blue arrows) superimposed on this background. Indexing these features yields a linear field dependence with a slope of $S = 262.6 \pm 3.2$~G [Fig.~\ref{fig:MR7K}(e)], consistent with AB oscillations arising from phase-coherent transport along the inner edge of the rings (see Table~\ref{tab:table1}). In a finer field sweep near zero [Fig.~\ref{fig:MR7K}(d)], the MR displays an additional set of oscillations with a dominant periodicity marked by red arrows. The corresponding field–index plot [Fig.~\ref{fig:MR7K}(f)] shows a linear slope of $S = 67.3 \pm 0.4$~G, consistent with LP oscillations due to fluxoid quantization from circulating bulk supercurrents, which confirms the formation of Cooper pairs and a macroscopic superconducting phase. A secondary set of oscillations with smaller periodicity (green arrows) also appears in the fine-field data [Figs.~\ref{fig:MR7K}(d,f)], corresponding to half-quantum vortices which are topological defects carrying half the superconducting flux quantum. Additional discussion of half-quantum vorticity is provided in Ref.~\cite{balakan2025observation}.

\begin{figure*}[t]
\includegraphics[width=0.99\textwidth]{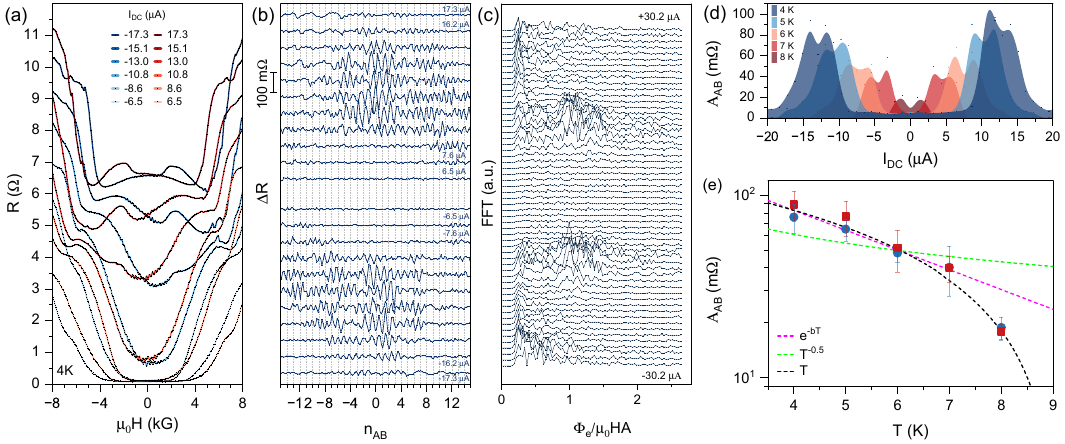}
\caption{\label{fig:MR4K} (a) Magnetoresistance measured at 4~K for various DC offsets, showing AB oscillations superimposed on a background governed by current--field symmetry $R(+I_\text{DC}, H) \approx R(-I_\text{DC}, -H)$. (b) Background-subtracted magnetoresistance $\Delta R$ plotted versus interference index. (c) Fast Fourier transform of $\Delta R$ at 4~K across a wide range of DC currents. 
(d) AB amplitude as a function of DC bias at temperatures from 4~K to 8~K. (e) Maximum amplitude of AB oscillations versus temperature. Dashed lines are fits to the data using the models described in the text.}
\end{figure*}

The coexistence of $h/e$ and $h/2e$ periodicities highlights the simultaneous presence of single-electron and Cooper-pair phase coherence in the same superconducting device. Fe(Te,Se) is generally regarded as a disordered material, with an electron mean free path on the order of 1~nm, characteristic of dirty metallic behavior \cite{Homes2015}. Given that the inner perimeter of each ring in our device is approximately $L = 1.6~\mu$m, it is unlikely for normal electrons to maintain phase coherence over such distances. Nonetheless, we observe robust $h/e$ oscillations, a hallmark of AB quantum interference from single electrons. Moreover, the magnetoresistance exhibits a clear current--field symmetry, $R(+I_\text{DC}, H) \approx R(-I_\text{DC}, -H)$, observable in both low- and high-field regimes at $T < T_c$ [Fig.~\ref{fig:MR4K}(a)]. This symmetry provides strong evidence for the presence of spin--orbit coupling under finite DC bias. Taken together, the interference periodicity, spatial confinement, robustness against disorder, and spin--orbit coupling provide compelling evidence that the observed AB oscillations originate from coherent edge states embedded within a superconducting background, consistent with helical or spin-polarized edge modes in a topological superconductor.

Aharonov--Bohm oscillations persist even at low temperatures. Figure~\ref{fig:MR4K}(a) shows magnetoresistance measured at 4~K for a range of DC offsets. At low bias ($|I_\text{DC}| < 7~\mu$A), the device remains fully superconducting at low fields, with vanishing resistance and a smooth parabolic background at higher fields due to the Meissner effect. At intermediate DC offsets, oscillations with a periodicity of $\sim250$~G emerge, indicative of AB interference. Simultaneously, MR deviates from a purely quadratic profile. Instead, magnetoresistance exhibits a clear current–field symmetry, characterized by $R(+I_\text{DC}, H) \approx R(-I_\text{DC}, -H)$. To isolate the oscillatory component, a polynomial background is subtracted from the raw MR traces over the $-3.5$ to $+3.5$~kG window. The resulting residual signal, $\Delta R = R - R_\mathrm{back}$, is plotted in Fig.~\ref{fig:MR4K}(b) as a function of the interference index $n_\mathrm{AB} = \mu_0 H / \Delta B_e^A$, where $\Delta B_e^A = \Phi_e / A$ [Table~\ref{tab:table1}]. AB oscillations are strongest in the range $I_\text{DC} \approx \pm(10$--$15)~\mu$A, with peak amplitude approaching 100~m$\Omega$. Outside this range, the oscillations rapidly diminish. A fast Fourier transform over a wide current range, shown in Fig.~\ref{fig:MR4K}(c), confirms that AB features appear only within a finite bias window centered near $\pm12~\mu$A, indicating a non-monotonic activation of coherent edge transport.

Figure~\ref{fig:MR4K}(d) summarizes the AB amplitude versus DC offset at several temperatures. The amplitude exhibits a symmetric bimodal profile, with clear peaks at both positive and negative bias. At 4~K, the maxima occur near $\pm12~\mu$A and shift toward $\pm2~\mu$A as temperature increases to 8~K. The oscillations are suppressed for both low and high $I_\text{DC}$, indicating that a finite current is required to enable phase-coherent edge conduction. The optimal DC offset corresponding to the maximum AB amplitude decreases with temperature at a rate of $2.9~\mu\mathrm{A}/\mathrm{K}$.

Interestingly, the maximum AB amplitude, shown in Fig.~\ref{fig:MR4K}(e), decreases \emph{linearly} with temperature and vanishes near $T_c=8~K$, coinciding with the onset of superconductivity. As a direct measure of quantum interference strength, the AB amplitude has long been used to probe decoherence in mesoscopic systems. In general, it follows $A_{AB} \propto e^{-L/L_\phi}$, where $L$ is the ring circumference and $L_\phi$ the phase coherence length~\cite{milliken1987}. At low temperatures, coherence is typically limited by inelastic scattering from electron--electron interactions or environmental coupling, following a dephasing rate $1/\tau_\phi \propto T$~\cite{seelig2001}. In diffusive systems, this gives rise to a temperature scaling of $L_\phi = \sqrt{D\tau_\phi} \propto T^{-0.5}$~\cite{appen1995, ludwig2004, russo2008}, whereas in ballistic systems $L_\phi = v_F \tau_\phi \propto T^{-1}$~\cite{hansen2001, dauber2017}, where $D$ and $v_F$ are diffusion coefficient and Fermi velocity, respectively. Edge states in the integer quantum Hall regime exhibit ballistic transport~\cite{roulleau2008}, while studies of Dirac fermions in topological insulators reveal AB amplitude scaling consistent with both diffusive regime, $A_{AB} \propto T^{-0.5}$~\cite{xiu2011, peng2010}, and ballistic transport, $A_{AB} \propto e^{-bT}$~\cite{dufouleur2013, wang2016}. However, neither model fully captures the behavior observed in Fig.~\ref{fig:MR4K}(e). While the overall temperature dependence of AB amplitude is linear, an exponential decay captures the experimental data for $T < 7$~K, indicating dominant ballistic transport in this temperature range (purple dashed line, $A_{AB} \propto e^{-bT}$ with $b \approx 0.25$~K$^{-1}$). Assuming $e^{-bT} = e^{-L/L_\phi}$, we estimate a coherence length of $L_\phi \approx 1.6~\mu$m at 4~K.

The suppression of AB amplitude near the superconducting transition can be understood through two possible mechanisms. One scenario is that, as the superconducting gap closes near $T_c$, topological protection is lost and the edge states vanish, reflecting bulk-boundary correspondence. Meanwhile, transport becomes dominated by incoherent bulk carriers, which obscure any residual phase-coherent contribution from the edge. While this picture is qualitatively consistent with the disappearance of interference, it does not account for the observed linear temperature dependence. Alternatively, due to the one-dimensional nature of edge states, the electron–electron interactions can strongly modify the single particle picture. We developed a theoretical model incorporating edge interactions that predicts a linear decay of AB amplitude with temperature, in agreement with experiment.

Helical or chiral edge states naturally emerge in 2D topological superconductors~\cite{read2000,zhang2021Sarma}. In contrast, proximitized topological surface states, as described by the Fu--Kane model, do not host edge modes as long as time-reversal symmetry is preserved. However, breaking time-reversal symmetry, for example by selectively populating one spin species~\cite{leggett2021}, allows a Zeeman field to induce edge states~\cite{fukane2008}. In the following, we outline a model based on the helical Luttinger liquid (HLL) theory~\cite{haldane1981,giamarchi2003,wu2006}, which describes interacting one-dimensional fermionic edge channels with spin-momentum locking (see Supplementary Material for details). 

Within HLL framework, the edge state is treated as a helical Luttinger liquid, where fermionic operators are expressed in terms of collective bosonic fields. The resulting correlation Green’s function,
$
    \mathcal{G}(x,t) = \langle \psi^\dagger(x,t) \psi(0,0)\rangle,
$
captures the phase coherence of electronic wavefunctions propagating around a ring threaded by magnetic flux which directly determines the amplitude of AB interference. The right- (left-)moving fermionic operator is given by
$
    \psi_{R(L)}(x,t) = (1/\sqrt{2\pi \zeta})e^{i[\varphi(x,t) \pm \theta(x,t)]},
$
where $\varphi(x,t)$ and $\theta(x,t)$ are Gaussian bosonic fields describing charge and current fluctuations, respectively, and $\zeta$ is a short-distance cutoff~\cite{von1998}. The correlation of bosonic fields for a full loop around the ring can be determined from finite-temperature field theory which leads to
$
\mathcal{G} = -\mathcal{A}_{AB}(T)\, e^{i 2\pi \phi / \phi_e},
$
where the minus sign reflects the nontrivial topology of the edge state~\cite{zhang2010}. $\mathcal{A}_{AB}$ represents the probability amplitude of AB interference
\begin{equation}
\label{eq:AB}
\mathcal{A}_{AB}(T) = \dfrac{1}{2 \pi \zeta} \left[ \frac{T_\zeta}{T} \sinh\left(\frac{T}{T_L}\right) \right]^{-\gamma}.
\end{equation}
Here, $T_x = \hbar \nu / \pi x k_B$ with $x=(\zeta,L)$ are characteristic temperature scales, and $\nu$ is the propagation velocity of bosonic fields. $\gamma = (K + K^{-1})/2 - 1$ is the interaction exponent, where $K < 1$ denotes the Luttinger parameter for repulsive interactions. It is noted that a Luttinger liquid with no interactions ($\gamma = 0$) exhibits a temperature-independent conductance~\cite{buttiker1986,kane1992}, and hence a constant AB amplitude. 

The characteristic temperature $T_L$ marks the crossover at which the AB amplitude begins to decrease with temperature. At very low temperatures, the amplitude becomes temperature-independent and saturates at $\mathcal{A}_{AB} = (1/2\pi \zeta)(\zeta/L)^\gamma$. At higher temperatures ($T > T_L$) and in the weakly correlated regime ($\gamma \ll 1$), Eq.~\ref{eq:AB} simplifies to
\begin{equation}
\label{eq:AB_Linear}
\mathcal{A}_{AB}(T) \approx b \left(1 - \gamma \frac{T}{T_L}\right),
\end{equation}
where $b$ is a constant. This linear dependence of the AB amplitude on temperature, characteristic of HLL edge states, is consistent with our experimental observation. A linear fit to the experimental data in Fig.~\ref{fig:MR4K}(e) gives a slope of $\gamma/T_L = 0.11$. In the weakly interacting limit, the propagation velocity of bosonic fields can be approximated by the Fermi velocity, $\nu \approx v_F \approx 3 \times 10^5$~m/s. This yields a characteristic crossover temperature of $T_L \approx 0.5$~K, and an interaction exponent of $\gamma \approx 0.05$. Using the relation $\gamma T / T_L = L / L_\phi$, we estimate a coherence length of $L_\phi \approx 3.6~\mu$m at 4~K within the HLL model.

The HLL model discussed above does not include inelastic dephasing due to spin-flip scattering from magnetic impurities. In the case of magnetic scattering, the AB oscillation amplitude increases with magnetic field, as sufficiently strong fields suppress spin-flip processes and restore phase coherence to the level set by electron–electron interactions~\cite{pierre2002}. Our measurements reveal no field-induced enhancement of the AB amplitude, suggesting that magnetic scattering does not play a significant role in the dephasing of edge states. It is known that bias current suppresses AB oscillations in a manner similar to temperature, primarily through energy averaging effects~\cite{appen1995,lin2010}. In contrast, our measurements reveal a non-monotonic dependence of AB amplitude on the DC offset, with edge state interference becoming active only within a finite bias range. The mechanism underlying this bias-dependent activation remains unclear and requires additional studies.

In summary, our results demonstrate that mesoscopic Fe(Te,Se) rings host coherent edge-state ballistic transport embedded within a superconducting background, consistent with the expected behavior of a topological superconductor. The bias-activated Aharonov--Bohm interference we observe opens a route toward edge-state interferometry in superconducting systems with nontrivial topology of order parameter and pave the way for future studies of Majorana zero modes and non-Abelian statistics.

\begin{acknowledgments}
%We wish to acknowledge...
\end{acknowledgments}

\bibliography{AB_EDGE_MAIN_REF}% Produces the bibliography via BibTeX.

\pagebreak
\widetext
\begin{center}
\textbf{\large Supplemental Materials}
\end{center}
%%%%%%%%%% Merge with supplemental materials %%%%%%%%%%
%%%%%%%%%% Prefix a "S" to all equations, figures, tables and reset the counter %%%%%%%%%%
\setcounter{equation}{0}
\setcounter{figure}{0}
\setcounter{table}{0}
\makeatletter
\renewcommand{\theequation}{S\arabic{equation}}
\renewcommand{\thefigure}{S\arabic{figure}}

\section{S1. AB interference within  helical Luttinger liquid theory}

We begin with the Hamiltonian of the helical Luttinger liquid (HLL) model~[\textcolor{blue}{1-3}], which describes interacting one-dimensional fermions with spin-momentum locking:
\begin{equation}
H = \frac{\hbar \nu}{2\pi} \int_0^L dx \left[ K (\partial_x \theta)^2 + \frac{1}{K} (\partial_x \varphi)^2 \right],
\end{equation}
where the bosonic fields satisfy the canonical commutation relation
$
[\varphi(x), \partial_y \theta(y)] = i \pi \delta(x - y).
$
Here, $\varphi(x)$ and $\theta(x)$ represent collective charge and current fluctuations, respectively, and serve as bosonic analogs of fermionic degrees of freedom. The parameter $\nu$ denotes the propagation velocity of bosonic excitations, and $L$ is the circumference of the ring. The Luttinger parameter $K$ characterizes the strength and nature of interactions: $K<1$ corresponds to repulsive interactions, $K>1$ to attractive ones, and $K=1$ to the non-interacting (free fermion) limit.

For a helical edge, the fermionic fields for right- and left-moving modes are bosonized as
\begin{equation}
\psi_R(x) = \frac{1}{\sqrt{2\pi \zeta}} e^{i[\varphi(x) + \theta(x)]}, \qquad 
\psi_L(x) = \frac{1}{\sqrt{2\pi \zeta}} e^{i[\varphi(x) - \theta(x)]},
\end{equation}
where $\zeta$ is a short-distance cutoff that regularizes the theory and ensures proper normalization. The prefactor $1/\sqrt{2\pi \zeta}$ accounts for the spatial smearing of fermionic operators intrinsic to bosonization. Due to spin-momentum locking, $\psi_R(x)$ describes right-moving (spin-up) electrons, while $\psi_L(x)$ corresponds to left-moving (spin-down) electrons.

A magnetic flux $\Phi$ threading the ring induces an Aharonov--Bohm (AB) phase for electrons traveling around the edge. This phase shift is given by
\begin{equation}
\delta \Phi_{\text{AB}} = \frac{2\pi \Phi}{\Phi_0},
\end{equation}
where $\Phi_0 = h/e$ is the flux quantum. This phase modifies the boundary condition of the edge state wavefunction. For AB interference, we focus on the coherence of a single species, for example right-moving (spin-up) electrons described by $\psi_R(x)$, as they traverse the ring. The AB oscillation amplitude is directly linked to the quantum coherence of an electron propagating around the ring, which is encoded in the correlation Green’s function:
\begin{equation}
\mathcal{G}(x,t) = \langle \psi^\dagger(x,t) \psi(0,0) \rangle.
\end{equation}
This function quantifies the amplitude for a fermion, introduced at position $0$ and time $0$, to be detected at position $x$ and time $t$, thus providing insight into how interactions and temperature affect coherence. Using the bosonized form of the fermionic operator for right-moving modes:
$
\psi_R(x,t) = \frac{1}{\sqrt{2\pi \zeta}}\, e^{i[\varphi(x,t) + \theta(x,t)]},
$
the Green’s function becomes
\begin{equation}
\mathcal{G}(x,t) = \frac{1}{2\pi \zeta} \left\langle e^{-i[\varphi(x,t)+\theta(x,t)]} e^{i[\varphi(0,0)+\theta(0,0)]} \right\rangle.
\end{equation}

Since $\varphi(x,t)$ and $\theta(x,t)$ are Gaussian bosonic fields, we can apply the identity
\begin{equation}
\langle e^{iA} e^{-iB} \rangle = e^{-\frac{1}{2} \langle (A - B)^2 \rangle},
\end{equation}
which yields
\begin{equation} \label{gf}
\mathcal{G}(x,t) = \frac{1}{2\pi \zeta} \exp\left[-\frac{1}{2} \left\langle \left( [\varphi(x,t) + \theta(x,t)] - [\varphi(0,0) + \theta(0,0)] \right)^2 \right\rangle \right].
\end{equation}

The next step involves evaluating the correlation functions
$\langle [\varphi(x,t) - \varphi(0,0)]^2 \rangle$ and $\langle [\theta(x,t) - \theta(0,0)]^2 \rangle$
at finite temperature. We then examine how magnetic flux modifies the behavior of the edge modes by altering the boundary conditions and zero-mode structure. The bosonic field $\varphi(x,t)$ on a ring of length $L$ at temperature $T$ can be mode-expanded as~[\textcolor{blue}{4}]:
\begin{equation} \label{field}
\varphi(x,t) = \varphi_0 + \frac{2\pi x}{L} N + \sum_{n\neq 0} \frac{1}{\sqrt{|n|}} \left( a_n e^{2\pi i n x / L} e^{-i \omega_n t} + \text{h.c.} \right).
\end{equation}
The first two terms represent the zero mode (\( n = 0 \)), with \( N \in \mathbb{Z} \) representing the winding number. The frequencies \( \omega_n = \frac{2\pi n}{\beta} \) are the bosonic Matsubara frequencies, where \( \beta = \frac{1}{k_B T} \), and \( a_n \), \( a_n^\dagger \) are the bosonic annihilation and creation operators, respectively.

In the presence of a magnetic flux \( \Phi \) threading the ring, the boundary condition for \( \varphi(x) \) becomes
\begin{equation}
\varphi(x+L) = \varphi(x) + 2\pi \frac{\Phi}{\Phi_0},
\end{equation}
which modifies the zero mode as
\begin{equation}
\varphi(x,t) = \varphi_0 + \frac{2\pi}{L} \left( N + \frac{\Phi}{\Phi_0} \right) x + \tilde{\varphi}(x,t).
\end{equation}

The oscillatory part is given by
\begin{equation}
\tilde{\varphi}(x,t) = \sum_{n\neq 0} \frac{1}{\sqrt{|n|}} \left( a_n e^{2\pi i n x / L} e^{-i \omega_n t} + \text{h.c.} \right),
\end{equation}
and satisfies periodic boundary conditions: $\tilde{\varphi}(x+L,t) = \tilde{\varphi}(x,t)$. These Fourier modes ($n \neq 0$) describe local particle–hole fluctuations and are insensitive to the topological phase accumulated due to magnetic flux. However, the zero-mode contribution
\begin{equation}
\varphi_{0}(x) = \frac{2\pi}{L} \left( N + \frac{\Phi}{\Phi_0} \right) x,
\end{equation}
solely determines the flux-dependent boundary condition. Consequently, magnetic flux enters the theory only through the zero-mode sector. As an electron encircles the ring, this term accumulates a total phase $2\pi \Phi/\Phi_0$, yielding the relation
\begin{equation}
e^{i[\varphi(L) - \varphi(0)]} = (-1)^N e^{-i 2\pi \Phi/\Phi_0}.
\end{equation}
In a 2D topological superconductor with a ring-shaped edge, the winding number $N = 1$ reflects the system’s nontrivial topology and encodes a Berry phase $\gamma = \pi$ associated with a spin-$1/2$ fermion encircling the flux~[\textcolor{blue}{1, 5}]. This topological phase appears explicitly as a prefactor in the fermionic Green’s function discussed in the next section. With the flux dependence clarified, we now examine how finite temperature impacts the coherence encoded in the Green’s function.

To understand how temperature influences coherence, we first derive the finite-temperature correlation functions of the bosonic fields $\varphi(x,t)$ and $\theta(x,t)$ that enter the Green’s function in Eq.~\ref{gf}. Since these fields are Gaussian, the correlators can be computed by thermal averaging over their normal mode expansions. The oscillatory component of the field is given by:
\begin{equation}
\tilde{\varphi}(x,t) = \sum_{q>0} \sqrt{\frac{2\pi}{qL}} \left[ a_q e^{i(qx - \omega_q t)} + a_q^\dagger e^{-i(qx - \omega_q t)} \right],
\end{equation}
where $\omega_q = v q$ with $q = 2\pi n / L$. The thermal average $\langle a_q^\dagger a_q \rangle = n_B(\omega_q)$ follows the Bose–Einstein distribution:
\begin{equation}
n_B(\omega_q) = \frac{1}{e^{\beta \hbar \omega_q} - 1}.
\end{equation}
The field correlation function then becomes:
\begin{equation}
\langle \varphi(x,t)\varphi(0,0) \rangle_T = \int_0^\infty \frac{dq}{q} \cos[q(x - \nu t)] \coth\left( \frac{\beta \hbar \nu q}{2} \right).
\end{equation}
This integral can be evaluated using contour integration or conformal mapping techniques, yielding:
\begin{equation}
\langle \varphi(x,t)\varphi(0,0) \rangle_T = -\frac{1}{4K} \ln \left[ \frac{ \sinh^2\left( \frac{\pi}{\beta \nu}(x + \nu t) \right) }{ \left( \frac{\pi \zeta}{\beta \nu} \right)^2 } \right],
\end{equation}
and similarly for $\theta(x,t)$:
\begin{equation}
\langle \theta(x,t)\theta(0,0) \rangle_T = -\frac{K}{4} \ln \left[ \frac{ \sinh^2\left( \frac{\pi}{\beta \nu}(x + \nu t) \right) }{ \left( \frac{\pi \zeta}{\beta \nu} \right)^2 } \right].
\end{equation}
The coefficients $K$ and $1/K$ reflect the strength of the interactions and their influence on the respective field fluctuations. A large $K$ implies stronger phase fluctuations (in $\theta$) and weaker density fluctuations (in $\varphi$). We neglect the mixed correlator \( \langle \varphi(x,t)\, \theta(0,0) \rangle \) and \( \langle \theta(x,t)\, \varphi(0,0) \rangle \) since it contributes only an overall phase to the Green’s function and does not affect the temperature-dependent amplitude of Aharonov–Bohm (AB) oscillations.

Using these temperature-dependent correlation functions, we evaluate the quantity entering Eq.~\ref{gf}:
\begin{equation}
\left\langle \left[ \varphi(x,t) + \theta(x,t) \right] \left[ \varphi(0,0) + \theta(0,0) \right] \right\rangle = -\left( \frac{1}{4K} + \frac{K}{4} \right) \ln \left[ \frac{ \sinh^2\left( \frac{\pi k_B T}{\hbar \nu} (x + \nu t) \right) }{ \left( \frac{\pi \zeta k_B T}{\hbar \nu} \right)^2 } \right].
\end{equation}
Defining the interaction parameter:
\begin{equation}
\gamma = \frac{1}{2} \left( K + \frac{1}{K} \right) - 1.
\end{equation}
The $-1$ ensures $\gamma = 0$ in the non-interacting limit ($K = 1$). This normalization isolates the contribution of electron–electron interactions to the suppression of coherence. So we can simplify the above as:
\begin{equation}
\left\langle \left[ \varphi(x,t) + \theta(x,t) \right] \left[ \varphi(0,0) + \theta(0,0) \right] \right\rangle = -\frac{\gamma}{2} \ln \left[ \frac{ \sinh\left( \frac{\pi k_B T}{\hbar \nu} (x + \nu t) \right) }{ \frac{\pi \zeta k_B T}{\hbar \nu} } \right].
\end{equation}
Thus, the finite-temperature Green’s function becomes:
\begin{equation}
\mathcal{G}(x,t) = \frac{1}{2\pi \zeta} \left[ \frac{ \frac{\pi \zeta k_B T}{\hbar \nu} }{ \sinh\left( \frac{\pi k_B T}{\hbar \nu} (x + \nu t) \right) } \right]^\gamma.
\end{equation}

To study the AB interference, we evaluate the correlation function around the full loop of the ring:
\begin{equation}
\mathcal{G}(L,T) = \frac{(-1)^N}{2\pi \zeta} \left( \frac{\frac{\pi \zeta k_B T}{\hbar \nu}}{\sinh\left( \frac{\pi L k_B T}{\hbar \nu} \right)} \right)^\gamma e^{i 2\pi \Phi / \Phi_0} = -\mathcal{A}_{AB}(T)\, e^{i 2\pi \Phi / \Phi_0},
\end{equation}
where the minus sign originates from the nontrivial topology of the system with winding number \( N = 1 \), and \( \mathcal{A}_{AB} \) denotes the temperature-dependent AB amplitude:
\begin{equation}
\mathcal{A}_{AB}(T) = \frac{1}{2\pi \zeta} \left( \frac{ \frac{\pi \zeta k_B T}{\hbar \nu} }{ \sinh\left( \frac{\pi L k_B T}{\hbar \nu} \right) } \right)^\gamma.
\end{equation}
In the non-interacting case (\( \gamma = 0 \)), the AB amplitude remains constant with temperature, consistent with theoretical predictions for non-interacting one-dimensional channels~[\textcolor{blue}{6, 7}].
The characteristic temperature scale,
\begin{equation}
T_L = \frac{\hbar \nu}{\pi k_B L},
\end{equation}
marks the crossover point beyond which dephasing becomes significant. 

In the low-temperature limit \( T < T_L \), we use the approximation \( \sinh(x) \approx x \) to obtain:

\begin{equation}
\mathcal{A}_{AB} \approx \frac{1}{2\pi \zeta} \left( \frac{\zeta}{L} \right)^\gamma,
\end{equation}

which indicates saturation of the AB amplitude and maximal coherence as \( T \to 0 \). Here, the dimensionless ratio \( \zeta/L \) encodes the scaling of microscopic interactions with system size, while the exponent \( \gamma \) controls the strength of interaction-induced dephasing.

In the higher temperature regime \( T > T_L \) and for weak interactions \( \gamma \ll 1 \), we expand \( \sinh^{-\gamma}(x) \) as:
\begin{equation}
\sinh^{-\gamma}(x) \approx 1 - \gamma(x - \log 2) + \mathcal{O}(\gamma^2),
\end{equation}
which captures leading-order interaction corrections. Substituting this into the amplitude expression yields:
\begin{equation}
\mathcal{A}_{AB} \approx \frac{1}{2\pi \zeta}  \left( \frac{T}{T_\zeta} \right)^\gamma \left[ 1 - \gamma \frac{T}{T_L} \right],
\end{equation}
where $T_\zeta = \hbar \nu/\pi k_B \zeta$. This result demonstrates the linear decay of the AB amplitude with increasing temperature, originating from interaction-induced dephasing.

\section{References}
\begingroup % Group the redefinition to limit its scope
  \renewcommand{\labelenumi}{[\theenumi]} % Redefine the label format
  \begin{enumerate}
    \item T. Giamarchi, Quantum physics in one dimension, Vol. 121 (Clarendon press, 2003).
    \item F. D. M. Haldane, Luttinger liquid theory of one-dimensional quantum fluids. I. properties of the Luttinger model and their extension to the general 1d interacting spinless fermi gas, Journal of Physics C: Solid State Physics 14, 2585 (1981).
    \item C. Wu, B. A. Bernevig, and S.-C. Zhang, Helical liquid and the edge of quantum spin hall systems, Phys. Rev. Lett. 96, 106401 (2006).
    \item J. Von Delft and H. Schoeller, Bosonization for beginners—refermionization for experts, Annalen der Physik 510, 225 (1998).
    \item Y. Zhang and A. Vishwanath, Anomalous Aharonov-Bohm conductance oscillations from topological insulator surface states, Phys. Rev. Lett. 105, 206601 (2010).
    \item M. B¨uttiker, Four-terminal phase-coherent conductance, Phys. Rev. Lett. 57, 1761 (1986).
    \item  C. L. Kane and M. P. A. Fisher, Transport in a one-channel Luttinger liquid, Phys. Rev. Lett.68, 1220 (1992).
  \end{enumerate}
\endgroup

\end{document}